\documentclass{IEEEtran}
\usepackage[T1]{fontenc}
\usepackage{float}
\usepackage{graphicx}

\makeatletter

\newcommand{\noun}[1]{\textsc{#1}}

 \newcommand{\lyxaddress}[1]{
   \par {\raggedright #1 
   \vspace{1.4em}
   \noindent\par}
 }

\makeatother
\begin{document}

\title{Analysis of Inter-Domain Traffic Correlations: Random Matrix Theory
Approach}

\author{Viktoria Rojkova, Mehmed Kantardzic}

\maketitle

\lyxaddress{Department of Computer Engineering and Computer Science, University
of Louisville, Louisville, KY 40292 email: \{vbrozh01, mmkant01\}@gwise.louisville.edu }

\begin{abstract}
The traffic behavior of University of Louisville network with the
interconnected backbone routers and the number of Virtual Local Area
Network (VLAN) subnets is investigated using the Random Matrix Theory
(RMT) approach. We employ the system of equal interval time series
of traffic counts at all router to router and router to subnet connections
as a representation of the inter-VLAN traffic. The cross-correlation
matrix $C$ of the traffic rate changes between different traffic
time series is calculated and tested against null-hypothesis of random
interactions. 

The majority of the eigenvalues $\lambda_{i}$ of matrix $C$ fall
within the bounds predicted by the RMT for the eigenvalues of random
correlation matrices. The distribution of eigenvalues and eigenvectors
outside of the RMT bounds displays prominent and systematic deviations
from the RMT predictions. Moreover, these deviations are stable in
time. 

The method we use provides a unique possibility to accomplish three
concurrent tasks of traffic analysis. The method verifies the uncongested
state of the network, by establishing the profile of random interactions.
It recognizes the system-specific large-scale interactions, by establishing
the profile of stable in time non-random interactions. Finally, by
looking into the eigenstatistics we are able to detect and allocate
anomalies of network traffic interactions.
\end{abstract}

\section*{Categories and Subject Descriptors}

C.2.3 {[}\textbf{Computer-Communication Networks}{]}: Network Operations

\section*{General Terms}

Measurement, Experimentation

\begin{keywords}
Network-Wide Traffic Analysis, Random Matrix Theory, Large-Scale Correlations 
\end{keywords}

\section{introduction}

The infrastructure, applications and protocols of the system of communicating
computers and networks are constantly evolving. The traffic, which
is an essence of the communication, presently is a voluminous data
generated on minute-by-minute basis within multi-layered structure
by different applications and according to different protocols. As
a consequence, there are two general approaches in analysis of the
traffic and in modeling of its healthy behavior. In the first approach,
the traffic analysis considers the protocols, applications, traffic
matrix and routing matrix estimates, independence of ingress and egress
points and much more. The second approach treats the infrastructure
between the points from which the traffic is obtained as a {}``black
box'' \cite{Lau,Allen}. 

Measuring interactions between logically and architecturally equivalent
substructures of the system is a natural extension of the {}``black
box'' approach. Certain amount of work in this direction has already
been done. Studies on statistical traffic flow properties revealed
the {}``congested'', {}``fluid'' and {}``transitional'' regimes
of the flow at a large scale \cite{Fukuda,Ohira}. The observed collective
behavior suggests the existence of the large-scale network-wide correlations
between the network subparts. Indeed, the \cite{Barthelemy} work
showed the large-scale cross-correlations between different connections
of the Renater scientific network. Moreover, the analysis of correlations
across all simultaneous network-wide traffic has been used in network
distributed attacks detection \cite{LCD}. 

The distributions and stability of established interactions statistics
represent the characteristic features of the system and may be exploited
in healthy network traffic profile creation, which is an essential
part of network anomaly detection. As it is successfully demonstrated
in  \cite{Crovella}, all tested traffic anomalies change the distribution
of the traffic features. 

Among numerous types of traffic monitoring variables, time series
of traffic counts are free of applications {}``semantics'' and thus
more preferable for {}``black box'' analysis. To extract the meaningful
information about underlying interactions contained in time series,
the empirical correlation matrix is a usual tool at hand. In addition,
there are various classes of statistical tools, such as principal
component analysis, singular value decomposition, and factor analysis,
which in turn strongly rely on the validity of the correlation matrix
and obtain the meaningful part of the time series. Thus, it is important
to understand quantitatively the effect of noise, i.e. to separate
the noisy, random interactions from meaningful ones. In addition,
it is crucial to consider the finiteness of the time series in the
determination of the empirical correlation, since the finite length
of time series available to estimate cross correlations introduces
{}``measurement noise'' \cite{Guhr1}. Statistically, it is also
advisable to develop null-hypothesis tests in order to check the degree
of statistical validity of the results obtained against cases of purely
random interactions.

The methodology of random matrix theory (RMT) developed for studying
the complex energy levels of heavy nuclei and is given a detailed
account in \cite{Wigner1,Dyson1,Dyson2,Mehta,Brody,Guhr3}. For our
purposes this methodology comes in as a series of statistical tests
run on the eigenvalues and eigenvectors of {}``system matrix'',
which in our case is traffic time series cross-correlation matrix
$C$ (and is Hamiltonian matrix in case of nuclei and other RMT systems
\cite{Wigner1,Dyson1,Dyson2,Mehta,Brody,Guhr3}).

In our study, we propose to investigate the network traffic as a complex
system with a certain degree of mutual interactions of its constituents,
i.e. single-link traffic time series, using the RMT approach. We concentrate
on the large scale correlations between the time series generated
by Simple Network Manage Protocol (SNMP) traffic counters at every
router-router and router-VLAN subnet connection of University of Louisville
backbone routers system.

The contributions of this study are as follows:

\begin{itemize}
\item We propose the application constraints free methodology of network-wide
traffic time series interactions analysis. Even though in this particular
study, we know in advance that VLANs represent separate broadcast
domains, VLAN-router incoming traffic is a traffic intended for other
VLANs and VLAN-router outgoing traffic is a routed traffic from other
VLANs. Nevertheless, this information is irrelevant for our analysis
and acquired only at the interpretation of the analysis results.
\item Using the RMT, we are able to separate the random interactions from
system specific interactions. The vast majority of traffic time series
interact in random fashion. The time stable random interactions signify
the healthy, and free of congestion traffic. The proposed analysis
of eigenvector distribution allows to verify the time series content
of uncongested traffic. 
\item The time stable non-random interactions provide us with information
about large-scale system-specific interactions.
\item Finally, the temporal changes in random and non-random interactions
can be detected and allocated with eigenvalues and eigenvectors statistics
of interactions.
\end{itemize}
The organization of this paper is as follows. Section II presents
the survey of related work. We describe the RMT methodology in Section
III. Section IV contains the explanation of the data analyzed. In
Section V we test the eigenvalue distribution of inter-VLAN traffic
time series cross-correlation matrix C against the RMT predictions.
In Section VI we analyze the content of inter-VLAN traffic interactions
by mean of eigenvalues and eigenvectors deviated from RMT. Section
VII discusses the characteristic traffic interactions parameters of
the system such as time stability of the deviating eigenvalues and
eigenvectors, inverse participation ratio (IPR) of eigenvalues spectra,
localization points in IPR plot, overlap matrices of the deviating
eigenvectors. With series of different experiments, we demonstrate
how traffic interactions anomalies can be detected and allocated in
time and space using various visualization techniques on eigenvalues
and eigenvectors statistics in Section VIII. We present our conclusions
and prospective research steps in Section IX.

\section{related work}

Few works investigate the interactions of traffic time series regardless
of underlying architecture of the traffic system. As it was stated
in Introduction, the study of \cite{Barthelemy} showed the large-scale
cross-correlations between different connections of the French scientific
network Renater with 26 interconnected routers and 650 connections
links. The random interactions between traffic time series of complex
traffic system without the routing protocol information were established
by Krbalek and Seba in \cite{Seba} for transportation system in Cuernavaca
(Mexico). 

The urgent need for a network-wide, scalable approach to the problem
of healthy network traffic profile creation is expressed in works
of \cite{Crovella,Crovella2,Min,McNutt,Roughan,Huang}. There are
several studies with the promising results, which demonstrate that
the traffic anomalous events cause the temporal changes in statistical
properties of traffic features. Lakhina, Crovella and Diot presented
the characterization of the network-wide anomalies of the traffic
flows. The authors studied three different types of traffic flows
and fused the information from flow measurements taken throughout
the entire network. They obtained and classified a different set of
anomalies for different traffic types using the subspace method \cite{Crovella2}.

The same group of researchers extended their work in \cite{Crovella}.
Under the new assumption that any network anomaly induces the changes
in distributional aspects of packet header fields, they detected and
identified large set of anomalies using the entropy measurement tool.

Hidden Markov model has been proposed to model the distribution of
network-wide traffic in \cite{Min}. The observation window is used
to distinguish denial of service (DoS) flooding attack mixed with
the normal background traffic. 

Roughan et al. combined the entire network routing and traffic data
to detect the IP forwarding anomalies \cite{Roughan}.

Huang et al., \cite{Huang} used the distributed version of the Principal
Component Analysis (PCA) method for centralized network-wide volume
anomaly detection. A key ingredient of their framework is an analytical
method based on stochastic matrix perturbation theory that balances
between the accuracy of the approximate network anomaly detection
and the amount of data communication over the network. 

The authors of \cite{McNutt} found the high temporal correlation
(frequently > 0.99) between flow counts on quiescent ports (TCP/IP
ports which are not in regular use) at the one of the known pre-attack,
so called \emph{reconnaissance}, anomalous behavior, vertical scan.

\section{rmt methodology}

The RMT was employed in the financial studies of stock correlations
\cite{Sharifi,Guhr1}, communication theory of wireless systems \cite{Tulino},
array signal processing \cite{Tse}, bioinformatics studies of protein
folding \cite{Zee}. We are not aware of any work, except for \cite{Barthelemy},
where RMT techniques were applied to the Internet traffic system. 

We adopt the methodology used in works on financial time series correlations
(see \cite{Sharifi,Guhr1} and references therein) and later in \cite{Barthelemy},
which discusses cross-correlations in Internet traffic. In particular,
we quantify correlations between $N$ traffic counts time series of
$L$ time points, by calculating the traffic rate change of every
time series $T$ $i=1,\dots,N$ , over a time scale $\Delta t$,\begin{equation}
G_{i}\left(t\right)\equiv\textrm{ln}\, T_{i}\left(t+\Delta t\right)-\textrm{ln}\, T_{i}\left(t\right)\label{eq1}\end{equation}
where $T{}_{i}\left(t\right)$ denotes the traffic rate of time series
$i$. This measure is independent from the volume of the traffic exchange
and allows capturing the subtle changes in the traffic rate \cite{Barthelemy}.
The normalized traffic rate change is

\begin{equation}
g_{i}\left(t\right)\equiv\frac{G_{i}\left(t\right)-\left\langle G_{i}\left(t\right)\right\rangle }{\sigma_{i}}\label{eq2}\end{equation}
where $\sigma_{i}\equiv\sqrt{\left\langle G_{i}^{2}\right\rangle -\left\langle G_{i}\right\rangle ^{2}}$
is the standard deviation of $G_{i}$. The equal-time cross-correlation
matrix $C$ can be computed as follows\begin{equation}
C_{ij}\equiv\left\langle g_{i}\left(t\right)g_{j}\left(t\right)\right\rangle \label{eq3}\end{equation}
The properties of the traffic interactions matrix $C$ have to be
compared with those of a random cross-correlation matrix \cite{Laloux}.
In matrix notation, the interaction matrix $C$ can be expressed as\begin{equation}
C=\frac{1}{L}GG^{T},\label{eq4}\end{equation}
where $G$ is $N\times L$ matrix with elements $\left\{ g_{i\, m}\equiv g_{i}\left(m\bigtriangleup t\right);\right.$
$i=1,\dots,N;$ $\left.m=0,\dots,L-1\right\} ,$ and $G^{T}$ denotes
the transpose of $G$. Just as was done in \cite{Guhr1}, we consider
a random correlation matrix \begin{equation}
R=\frac{1}{L}AA^{T},\label{eq5}\end{equation}
where $A$ is $N\times L$ matrix containing $N$ time series of $L$
random elements $a_{i\, m}$ with zero mean and unit variance, which
are mutually uncorrelated as a null hypothesis.

Statistical properties of the random matrices $R$ have been known
for years in physics literature \cite{Wigner1,Brody,Dyson1,Dyson2,Mehta,Guhr3}.
In particular, it was shown analytically \cite{Sengupta} that, under
the restriction of $N\rightarrow\infty,$ $L\rightarrow\infty$ and
providing that $Q\equiv L/N$$\left(>1\right)$ is fixed, the probability
density function $P_{rm}\left(\lambda\right)$ of eigenvalues $\lambda$
of the random matrix $R$ is given by

\begin{equation}
P_{rm}\left(\lambda\right)=\frac{Q}{2\pi}\frac{\sqrt{\left(\lambda_{+}-\lambda\right)\left(\lambda-\lambda_{-}\right)}}{\lambda}\label{eq6}\end{equation}
where $\lambda_{+}$ and $\lambda_{-}$ are maximum and minimum eigenvalues
of $R,$ respectively and $\lambda_{-}\leq\lambda_{i}\leq\lambda_{+}$.
$\lambda_{+}$ and $\lambda_{-}$are given analytically by

\begin{equation}
\lambda_{\pm}=1+\frac{1}{Q}\pm2\sqrt{\frac{1}{Q}}.\label{eq7}\end{equation}
Random matrices display \emph{universal} functional forms for eigenvalues
correlations which depend on the general symmetries of the matrix
only. First step to test the data for such a universal properties
is to find a transformation called {}``unfolding'', which maps the
eigenvalues $\lambda_{i}$ to new variables, {}``unfolded eigenvalues''
$\xi_{i},$ whose distribution is uniform \cite{Mehta,Brody,Guhr3}.
Unfolding ensures that the distances between eigenvalues are expressed
in units of \emph{local} mean eigenvalues spacing \cite{Mehta}, and
thus facilitates the comparison with analytical results.

We define the cumulative distribution function of eigenvalues, which
counts the number of eigenvalues in the interval $\lambda_{i}\leq\lambda,$

\begin{equation}
F\left(\lambda\right)=N\int_{-\infty}^{\lambda}P\left(x\right)dx,\label{eq8}\end{equation}
where $P\left(x\right)$ denotes the probability density of eigenvalues
and $N$ is the total number of eigenvalues. The function $F\left(\lambda\right)$
can be decomposed into an average and a fluctuating part, \begin{equation}
F\left(\lambda\right)=F_{av}\left(\lambda\right)+F_{fluc}\left(\lambda\right),\label{eq9}\end{equation}
Since $P_{fluc}\equiv dF_{fluc}\left(\lambda\right)/d\lambda=0$ on
average, \begin{equation}
P_{rm}\left(\lambda\right)\equiv\frac{dF_{av}\left(\lambda\right)}{d\lambda},\label{eq10}\end{equation}
is the averaged eigenvalues density. The dimensionless, unfolded eigenvalues
are then given by \begin{equation}
\xi_{i}\equiv F_{av}\left(\lambda_{i}\right).\label{eq11}\end{equation}

Three known universal properties of GOE matrices (matrices whose elements
are distributed according to a Gaussian probability measure) are:
(i) the distribution of nearest-neighbor eigenvalues spacing $P_{GOE}\left(s\right)$
\begin{equation}
P_{GOE}\left(s\right)=\frac{\pi s}{2}exp\left(-\frac{\pi}{4}s^{2}\right),\label{eq12}\end{equation}
(ii) the distribution of next-nearest-neighbor eigenvalues spacing,
which is according to the theorem due to \cite{Dyson2} is identical
to the distribution of nearest-neighbor spacing of Gaussian symplectic
ensemble (GSE),

\begin{equation}
P_{GSE}\left(s\right)=\frac{2^{18}}{3^{6}\pi^{3}}s^{4}exp\left(-\frac{64}{9\pi}s^{2}\right)\label{eq13}\end{equation}
and finally (iii) the {}``number variance'' statistics $\Sigma^{2}$,
defined as the variance of the number of unfolded eigenvalues in the
intervals of length $l$, around each $\xi_{i}$ \cite{Mehta,Guhr3,Brody}.\begin{equation}
\Sigma^{2}\left(l\right)=\left\langle \left[n\left(\xi,l\right)-l\right]^{2}\right\rangle _{\xi},\label{eq14}\end{equation}
where $n\left(\xi,l\right)$ is the number of the unfolded eigenvalues
in the interval $\left[\xi-\frac{l}{2},\xi+\frac{l}{2}\right]$. The
number variance is expressed as follows

\begin{equation}
\Sigma^{2}\left(l\right)=l-2\int_{0}^{l}\left(l-x\right)Y\left(x\right)dx,\label{eq15}\end{equation}
where $Y\left(x\right)$ for the GOE case is given by \cite{Mehta}

\begin{equation}
Y\left(x\right)=s^{2}\left(x\right)+\frac{ds}{dx}\int_{x}^{\infty}s\left(x'\right)dx',\label{eq16}\end{equation}
and \begin{equation}
s\left(x\right)=\frac{sin\left(\pi x\right)}{\pi x}.\label{eq17}\end{equation}
Just as was stressed in \cite{Guhr1,Sharifi,Stockman} the overall
time of observation is crucial for explaining the empirical cross-correlation
coefficients. On one hand, the longer we observe the traffic the more
information about the correlations we obtain and less {}``noise''
we introduce. On the other hand, the correlations are not stationary,
i.e. they can change with time. To differentiate the {}``random''
contribution to empirical correlation coefficients from {}``genuine''
contribution, the eigenvalues statistics of $C$ is contrasted with
the eigenvalues statistics of a correlation matrix taken from the
so called {}``chiral'' Gaussian Orthogonal Ensemble \cite{Guhr1}.
Such an ensemble is one of the ensembles of RMT \cite{Stockman,Bouchaud},
briefly discussed in Appendix A. A \emph{random} cross-correlation
matrix, which is a matrix filled with uncorrelated Gaussian random
numbers, is supposed to represent transient uncorrelated in time network
activity, that is, a completely noisy environment. In case the cross-correlation
matrix $C$ obeys the same eigenstatistical properties as the RMT-matrix,
the network traffic is equilibrated and deemed universal in a sense
that every single connection interacts with the rest in a completely
chaotic manner. It also means a complete absence of congestions and
anomalies. Meantime, any stable in time deviations from the \emph{universal}
predictions of RMT signify system-specific, nonrandom properties of
the system, providing the clues about the nature of the underlying
interactions. That allows us to establish the profile of system-specific
correlations.

\section{data}

In this paper, we study the averaged traffic count data collected
from all router-router and router-VLAN subnet connections of the University
of Louisville backbone routers system. The system consists of nine
interconnected multi-gigabit backbone routers, over $200$ Ethernet
segments and over $300$ VLAN subnets. We collected the traffic count
data for $3$ months, for the period from September \emph{$21$, $2006$}
to December \emph{$20$, $2006$} from $7$ routers, since two routers
are reserved for server farms. The overall data amounted to approximately
$18$ GB.

The traffic count data is provided by Multi Router Traffic Grapher
(MRTG) tool that reads the SNMP traffic counters. MRTG log file never
grows in size due to the data consolidation algorithm: it contains
records of average incoming, outgoing, max and min transfer rate in
bytes per second with time intervals $300$ seconds, \emph{$30$}
minutes, $1$ day and $1$ month. We extracted $300$ seconds interval
data for seven days. Then, we separated the incoming and outgoing
traffic counts time series and considered them as independent. For
$352$ connections we formed $L=2015$ records of $N=704$ time series
with $300$ seconds interval. 

We pursued the changes in the traffic rate, thus, we excluded from
consideration the connections, where channel is open but the traffic
is not established or there is just constant rate and equal low amount
test traffic. Another reason for excluding the {}``empty'' traffic
time series is that they make the time series cross-correlation matrix
unnecessary sparse. The exclusion does not influence the analysis
and results. After the exclusions the number of the traffic time series
became $N=497$. 

To calculate the traffic rate change $G_{i}\left(t\right)$ we used
the logarithm of the ratio of two successive counts. As it is stated
earlier, $log$-transformation makes the ratio independent from the
traffic volume and allows capturing the subtle changes in the traffic
rate. We added 1 byte to all data points, to avoid manipulations with
$log\left(0\right)$, in cases where traffic count is equal to zero
bytes. This measure did not affect the changes in the traffic rate.

\section{eigenvalue distribution of cross-correlation matrix, comparison with
rmt}

We constructed inter-VLAN traffic cross-correlation matrix $C$ with
number of time series $N=497$ and number of observations per series
$L=2015$, ($Q=4.0625$) so that, $\lambda_{+}=2.23843$ and $\lambda_{-}=0.253876$.
Our first goal is to compare the eigenvalue distribution $P\left(\lambda\right)$
of $C$ with $P_{rm}\left(\lambda\right)$ \cite{Laloux}. To compute
eigenvalues of $C$ we used standard \emph{MATLAB} function. The empirical
probability distribution $P\left(\lambda\right)$ is then given by
the corresponding histogram. We display the resulting distribution
$P\left(\lambda\right)$ in Figure 1 and compare it to the probability
distribution $P_{rm}\left(\lambda\right)$ taken from Eq. (\ref{eq6})
calculated for the same value of traffic time series parameters ($Q=4.0625$).
The solid curve demonstrates $P_{rm}\left(\lambda\right)$ of Eq.(\ref{eq6}).
The largest eigenvalue shown in inset has the value $\lambda_{497}=8.99$.
We zoom in the deviations from the RMT predictions on the inset to
Figure 1. %
\begin{figure}[H]
\begin{center}\includegraphics{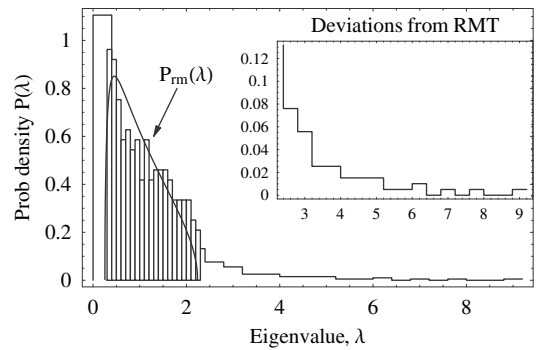}\end{center}

\caption{\label{1} Empirical probability distribution function $P\left(\lambda\right)$
for the inter-VLAN traffic cross-correlations matrix $C$ (histogram). }
\end{figure}
We note the presence of {}``bulk'' (RMT-like) eigenvalues which
fall within the bounds {[}$\lambda_{-},$$\lambda_{+}${]} for $P_{rm}\left(\lambda\right)$,
and presence of the eigenvalues which lie outside of the {}``bulk'',
representing deviations from the RMT predictions. In particular, largest
eigenvalue $\lambda_{497}=8.99$ for seven days period is approximately
four times larger than the RMT upper bound $\lambda_{+}$.

The histogram for well-defined bulk agrees with $P_{rm}\left(\lambda\right)$
suggesting that the cross-correlations of matrix $C$ are mostly random.
We observe that inter-VLAN traffic time series interact mostly in
a random fashion.

Nevertheless, the agreement of empirical probability distribution
$P\left(\lambda\right)$ of the bulk with $P_{rm}\left(\lambda\right)$
is not sufficient to claim that the bulk of eigenvalue spectrum is
random. Therefore, further RMT tests are needed \cite{Guhr1}.

To do that, we obtained the unfolded eigenvalues $\xi_{i}$ by following
the phenomenological procedure referred to as Gaussian broadening
\cite{Bruus}, (see \cite{Bruus,Bruus2,Guhr1,Sharifi}). The empirical
cumulative distribution function of eigenvalues $F\left(\lambda\right)$
agrees well with the $F_{av}\left(\lambda\right)$ (see Figure 2),
where $\xi_{i}$ obtained with Gaussian broadening procedure with
the broadening parameter $a=8$.%
\begin{figure}[h]
\begin{center}\includegraphics{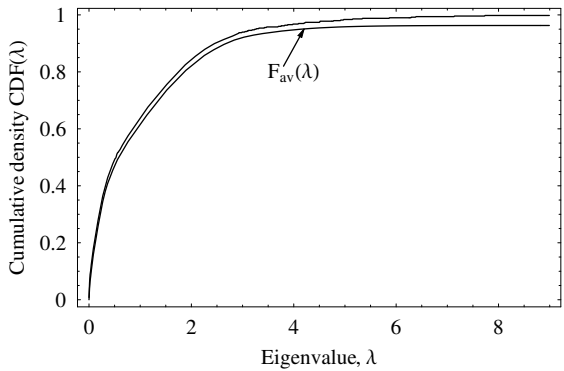}\end{center}

\caption{\label{2}The empirical cumulative distribution of $\lambda_{i}$
and unfolded eigenvalues $\xi_{i}\equiv F_{av}\left(\lambda\right)$. }
\end{figure}
The first independent RMT test is the comparison of the distribution
of the nearest-neighbor unfolded eigenvalue spacing $P_{nn}\left(s\right)$,
where $s\equiv\xi_{k+1}-\xi_{k}$ with $P_{GOE}\left(s\right)$ \cite{Mehta,Brody,Guhr3}.
The empirical probability distribution of nearest-neighbor unfolded
eigenvalues spacing $P_{nn}\left(s\right)$ and $P_{GOE}\left(s\right)$
are presented in Figure 3. The Gaussian decay of $P_{GOE}\left(s\right)$
for large $s$ suggests that $P_{GOE}\left(s\right)$ {}``probes''
scales only of the order of one eigenvalue spacing. The solid line
represents.%
\begin{figure}[h]
\begin{center}\includegraphics{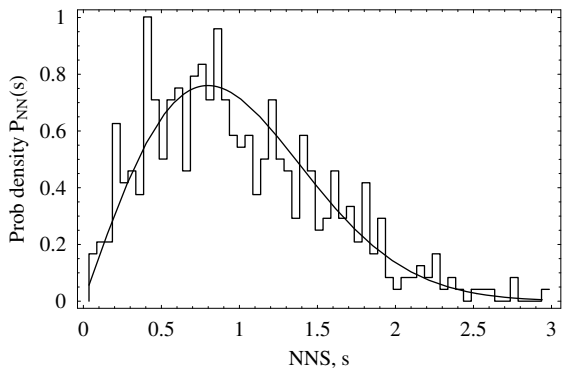}\end{center}

\caption{\label{3} Nearest-neighbor spacing distribution $P_{nn}\left(s\right)$
of unfolded eigenvalues $\xi_{i}$ of cross-correlation matrix $C$. }
\end{figure}
The agreement between empirical probability distribution $P_{nn}\left(s\right)$
and the distribution of nearest-neighbor eigenvalues spacing of the
GOE matrices $P_{GOE}\left(s\right)$ testifies that the positions
of two adjacent empirical unfolded eigenvalues at the distance $s$
are correlated just as the eigenvalues of the GOE matrices.

Next, we took on the distribution $P_{nnn}\left(s'\right)$ of next-nearest-neighbor
spacings $s'\equiv\xi_{k+2}-\xi_{k}$ between the unfolded eigenvalues.
According to \cite{Dyson2} this distribution should fit to the distribution
of nearest-neighbor spacing of the GSE. We demonstrate this correspondence
in Figure 4. The solid line shows $P_{GSE}\left(s\right)$.%
\begin{figure}[h]
\begin{center}\includegraphics{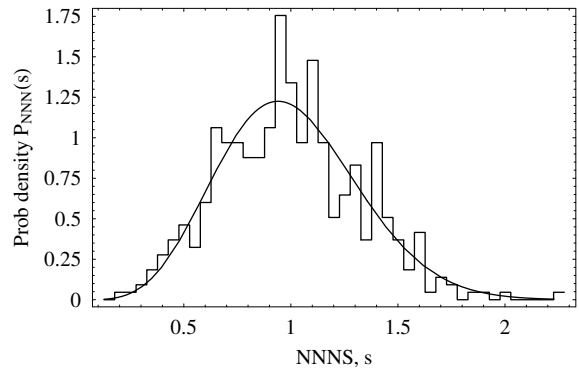}\end{center}

\caption{\label{4} Next-nearest-neighbor eigenvalue spacing distribution
$P_{nnn}\left(s'\right).$ }
\end{figure}
Finally, the long-range two-point eigenvalue correlations were tested.
It is known \cite{Mehta,Brody,Guhr3}, that if eigenvalues are uncorrelated
we expect the number variance to scale with $l$, $\Sigma^{2}\sim l$.
Meanwhile, when the unfolded eigenvalues of $C$ are correlated, $\Sigma^{2}$
approaches constant value, revealing {}``spectral rigidity'' \cite{Mehta,Brody,Guhr3}.
In Figure 5, we contrasted Poissonian number variance with the one
we observed, and came to the conclusion that eigenvalues belonging
to the {}``bulk'' clearly exhibit universal RMT properties. The
broadening parameter $a=8$ was used in Gaussian broadening procedure
to unfold the eigenvalues $\lambda_{i}$ \cite{Bruus,Bruus2,Guhr1,Sharifi}.
The dashed line corresponds to the case of uncorrelated eigenvalues.%
\begin{figure}[h]
\begin{center}\includegraphics{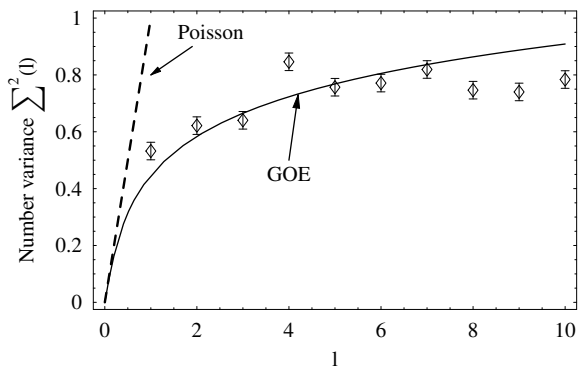}\end{center}

\caption{\label{5} Number variance $\Sigma^{2}\left(l\right)$ calculated
from the unfolded eigenvalues $\xi_{i}$ of $C$. }
\end{figure}
These findings show that the system of inter-VLAN traffic has a \emph{universal}
part of eigenvalues spectral correlations, shared by broad class of
systems, including chaotic and disordered systems, nuclei, atoms and
molecules. Thus it can be concluded, that the bulk eigenvalue statistics
of the inter-VLAN traffic cross-correlation matrix $C$ are consistent
with those of real symmetric random matrix $R$, given by Eq. (\ref{eq5})
\cite{Sengupta}. Meantime, the deviations from the RMT contain the
information about the system-specific correlations. The next section
is entirely devoted to the analysis of the eigenvalues and eigenvectors
deviating from the RMT, which signifies the meaningful inter-VLAN
traffic interactions.

\section{inter-vlan traffic interactions analysis}

We overview the points of interest in eigenvectors of inter-VLAN traffic
cross-correlation matrix $C$, which are determined according to $Cu^{k}=\lambda_{k}u^{k}$,
where $\lambda_{k}$ is $k$-th eigenvalue. Particularly important
characteristics of eigenvectors, proven to be useful in physics of
disordered conductors is the inverse participation ratio (IPR) (see,
for example, Ref. \cite{Guhr3}). In such systems, the IPR being a
function of an eigenstate (eigenvector) allows to judge and clarify
whether the corresponding eigenstate, and therefore electron is extended
or localized.

\subsection{Inverse participation ratio of eigenvectors components}

For our purposes, it is sufficient to know that IPR quantifies the
reciprocal of the number of significant components of the eigenvector.
For the eigenvector $u^{k}$ it is defined as\begin{equation}
I^{k}\equiv\sum_{l=1}^{N}\left[u_{l}^{k}\right]^{4},\label{eq18}\end{equation}
where $u_{l}^{k}$, $l=1,\dots,497$ are components of the eigenvector
$u^{k}$. In particular, the vector with one significant component
has $I^{k}=1$, while vector with identical components $u_{l}^{k}=1/\sqrt{N}$
has $I^{k}=1/N$.%
\begin{figure}[h]
\begin{center}\includegraphics{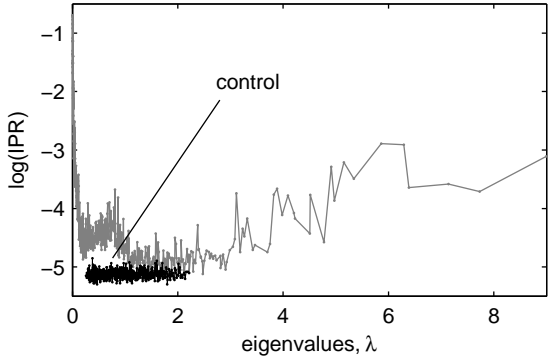}\end{center}

\caption{\label{6} Inverse participation ratio as a function of eigenvalue
$\lambda$.}
\end{figure}
Consequently, the inverse of IPR gives us a number of significant
participants of the eigenvector. In Figure 6 we plot the IPR of cross-correlation
matrix $C$ as a function of eigenvalue $\lambda$. The control plot
is IPR of eigenvectors of random cross-correlation matrix $R$ of
Eq. \ref{eq5}. As we can see, eigenvectors corresponding to eigenvalues
from $0.25$ to $3.5$, what is within the RMT boundaries, have IPR
close to $0$. This means that almost all components of eigenvectors
in the bulk interact in a random fashion. The number of significant
components of eigenvectors deviating from the RMT is typically twenty
times smaller than the one of the eigenvectors within the RMT boundaries,
around twenty. For instance, IPR of eigenvector $u^{492}$, which
corresponds to the eigenvalue $5.9$ in Figure 6, is $0.05$, i.e.
twenty time series are significantly contribute to $u^{492}$. Another
observation which we derive from Figure 6 is that the number of eigenvectors
significant participants is considerably smaller at both edges of
the eigenvalue spectrum. These findings resemble the results of \cite{Guhr1},
where the eigenvectors with a few participating components were referred
to as \emph{localized} vectors. The theory of \emph{localization}
is explained in the context of random band matrices, where elements
independently drawn from different probability distributions \cite{Guhr1}.
These matrices despite their randomness, still contain probabilistic
information. The \emph{localization} in inter-VLAN traffic is explained
as follows. The separated broadcast domains, i.e. VLANs forward traffic
from one to another only through the router, reducing the routing
for broadcast containment. Although the optimal VLAN deployment is
to keep as much traffic as possible from traversing through the router,
the bottleneck at the large number of VLANs is unavoidable.

\subsection{Distribution of eigenvectors components}

Another target of interest is the distribution of the components $\left\{ u_{l}^{k};\, l=1,\dots,N\right\} $
of eigenvector $u^{k}$ of the interactions matrix $C$. To calculate
vectors $u$ we used the \emph{MATLAB} routine again and obtained
components distribution $p\left(u\right)$ of the eigenvectors components.
Then, we contrasted it with the RMT predictions for the eigenvector
distribution $p_{rm}\left(u\right)$ of the random correlation matrix
$R$. According to \cite{Guhr3} $p_{rm}\left(u\right)$ has a Gaussian
distribution with mean zero and unit variance, i.e.\begin{equation}
p_{rm}\left(u\right)=\frac{1}{\sqrt{2\pi}}exp\left(\frac{-u^{2}}{2}\right).\label{eq19}\end{equation}
The weights of randomly interacting traffic counts time series, which
are represented by the eigenvectors components has to be distributed
normally. The results are presented in Figure 7. One can see (from
Figures 7a and 7b) that $p\left(u\right)$ for two $u^{k}$ taken
from the bulk is in accord with $p_{rm}\left(u\right)$. The distribution
$p\left(u\right)$ corresponding to the eigenvalue $\lambda_{i}$,
which exceeds the RMT upper bound ($\lambda_{i}>\lambda_{+}$), is
shown in Figure 7c. The solid line shows $p_{rm}\left(u\right)$ from
Eq. \ref{eq19}. (c) $p\left(u\right)$ for $u^{496}$, corresponding
to the eigenvalue outside of the RMT bulk. (d) $p\left(u\right)$
for $u^{497}$, corresponding to largest eigenvalue.%
\begin{figure}[H]
\begin{center}\includegraphics{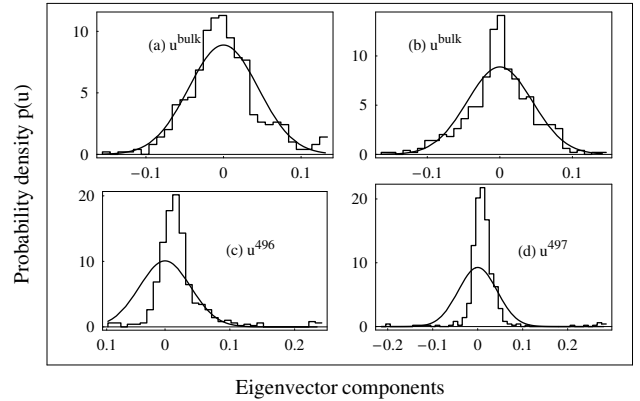}\end{center}

\caption{\label{7} Distribution of components $p\left(u\right)$ of eigenvectors
corresponding to eigenvalues (a) from the middle of the bulk, i.e.$\lambda_{-}<\lambda<\lambda_{+}$,
(b) from the bulk close to $\lambda_{+}$, (c) $\lambda_{496}$ (d)
$\lambda_{497}$.}
\end{figure}

\subsection{Deviating eigenvalues and significant inter-VLAN traffic series contributing
to the deviating eigenvectors.}

The distribution of $u^{497}$, the eigenvector corresponding to the
largest eigenvalue $\lambda_{497}$, deviates significantly from the
Gaussian (as follows from Figure 7d). While Gaussian kurtosis has
the value 3, the kurtosis of $p\left(u^{497}\right)$ comes out to
$23.22$. The smaller number of significant components of the eigenvector
also influences the difference between Gaussian distribution and empirical
distribution of eigenvector components. More than half of $u^{497}$components
have the same sign, thus slightly shifting the $p\left(u\right)$
to one side. This result suggests the existence of the common VLAN
traffic intended for inter-VLAN communication that affects all of
the significant participants of the eigenvector $u^{497}$with the
same bias. We know that the number of significant components of $u^{497}$
is twenty two, since IPR of $u^{497}$is $0.045$. Hence, the largest
eigenvector content reveals 22 traffic time series, which are affected
by the same event. We obtain the time series, which affects 22 traffic
time series by the following procedure. First of all, we calculate
projection $G^{497}\left(t\right)$ of the time series $G_{i}\left(t\right)$
on the eigenvector $u^{497}$,\begin{equation}
G^{497}\left(t\right)\equiv\sum_{i=1}^{497}u_{i}^{497}G_{i}\left(t\right)\label{eq20}\end{equation}
Next, we compare $G^{497}\left(t\right)$ with $G_{i}\left(t\right)$,
by finding the correlation coefficient $\left\langle \frac{G^{497}\left(t\right)}{\sigma^{497}}\frac{G_{i}\left(t\right)}{\sigma_{i}}\right\rangle $.
The Fiber Distributed Data Interface (FDDI)-VLAN internet switch at
one of the routers demonstrates the largest correlation coefficient
of $0.89$ (see Figure 8).%
\begin{figure}[h]
\begin{center}\includegraphics{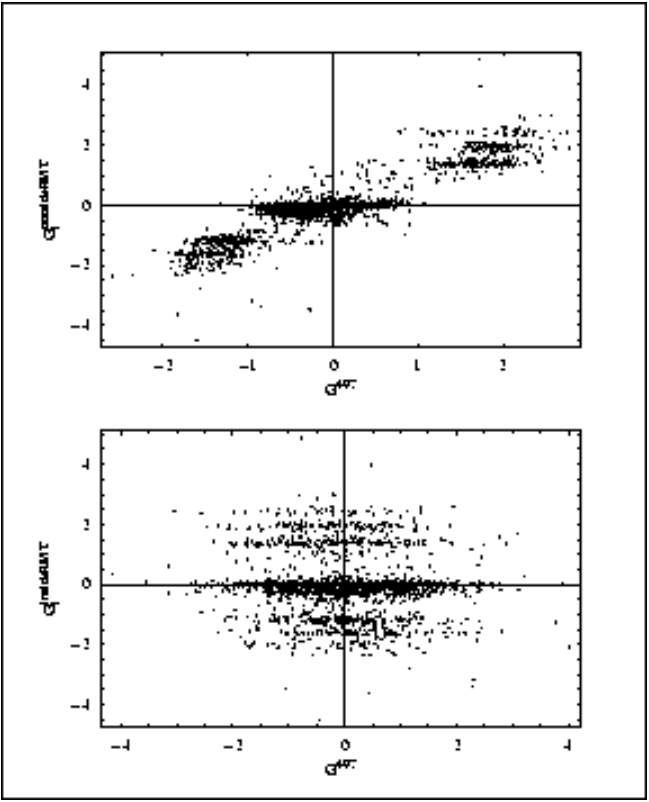}\end{center}

\caption{\label{8} (a) FDDI-VLAN internet switch time series regressed against
the projection $G^{497}\left(t\right)$ from Eq. \ref{eq20}. (b)
Time series defined by the eigenvector corresponding to eigenvalue
within RMT bounds shows no linear dependence on $G^{497}\left(t\right).$}
\end{figure}
The eigenvector $u^{497}$ has the following content: seven most significant
participants are seven FDDI-VLAN switches at the seven routers. The
presence of FDDI-VLAN switch provide us with information about VLAN
membership definition. FDDI is layer 2 protocol, which means that
at least one of two layer 2 membership is used, port group or/and
MAC address membership. The next group of significant participants
comprises of VLAN traffic intended for routing and already routed
traffic from different VLANs. The final group of significant participants
constitutes open switches, which pick up any {}``leaking'' traffic
on the router. Usually, the {}``leaking'' traffic is the network
management traffic, a very low level traffic which spikes when queried
by the management systems.

If every deviating eigenvalue notifies a particular sub-model of non-random
interactions of the network, then every corresponding eigenvector
presents the number of significant dimensions of sub-model. Thus,
we can think of every deviating eigenvector as a representative network-wide
{}``snapshot'' of interactions within the certain dimensions. 

The analysis of the significant participants of the deviating eigenvectors
revealed three types of inter-VLAN traffic time series groupings.
One group contains time series, which are interlinked on the router.
We recognize them as, router1-VLAN\_1000 traffic, router1-firewall
traffic and VLAN\_1000-router1 traffic. The time series, which are
listed as router1-vlan\_2000, router2-VLAN\_2000, router3-VLAN\_2000,
etc., are reserved for the same service VLAN on every router and comprise
another group. The content of these groups suggests the VLANs implementation,
it is a mixture of infrastructural approach, where functional groups
(departments, schools, etc.) are considered, and service approach,
where VLAN provides a particular service (network management, firewall,
etc.).

\section{stability of inter-vlan traffic interactions in time}

We expect to observe the stability of inter-VLAN traffic interactions
in the period of time used to compute traffic cross-correlation matrix
$C$. The eigenvalues distribution at different time periods provides
the information about the system stabilization, i.e. about the time
after which the fluctuations of eigenvalues are not significant. Time
periods of $1$ hour, $3$ hours and $6$ hours are not sufficient
to gain the knowledge about the system, which is demonstrated in Figure
9a. In Figure 9b the system stabilizes after $1$ day period. To observe
the time stability of inter-VLAN meaningful interactions we computed
the {}``overlap matrix'' of the deviating eigenvectors for the time
period $t$ and deviating eigenvectors for the time period $t+\tau$,
where $t=60h,\tau=\left\{ 0h,3h,12h,24h,36h,48h\right\} $.%
\begin{figure}[h]
\begin{center}\includegraphics{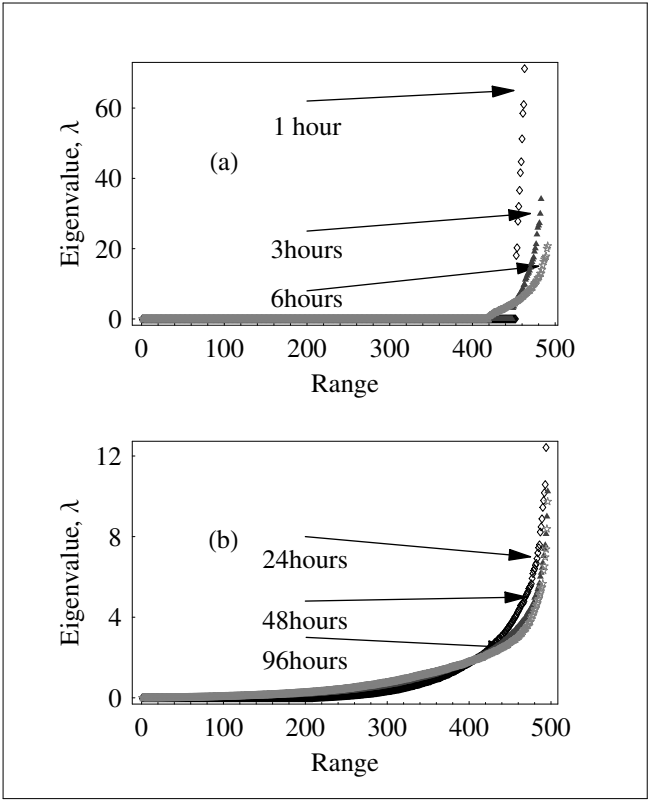}\end{center}

\caption{\label{9} (a) Eigenvalues distributions of traffic streams correlation
matrix $C$ for $1$ hour, $3$ hours and $6$ hours time intervals.
(b) Eigenvalues distributions for $24$ hours, $48$ hours and $72$
hours}
\end{figure}

First, we obtained matrix D from $p=57$ eigenvectors, which correspond
to $p$ eigenvalues outside of the RMT upper bound $\lambda_{+}$.
Then we computed the {}``overlap matrix'' $O\left(t,\tau\right)$
from $D_{A}D_{B}^{T}$, where $O_{ij}$ is a scalar product of the
eigenvector $u^{i}$ of period $A$ (starting at time $t=t$) with
$\textrm{u}^{j}$ of period $B$ at the time $t=t+\tau$,

\begin{equation}
O_{ij}\left(t,\tau\right)\equiv\sum_{k=1}^{N}D_{ik}\left(t\right)D_{ik}\left(t+\tau\right)\label{eq21}\end{equation}
The values of $O_{ij}\left(t,\tau\right)$ elements at $i=j$, i.e.
of diagonal elements of matrix $O$ will be $1$, if the matrix $D\left(t+\tau\right)$
is identical to the matrix $D\left(t\right)$. Clearly, the diagonal
of the {}``overlap matrix'' $O$ can serve as an indicator of time
stability of $p$ eigenvectors outside of the RMT upper bound $\lambda_{+}$.
The gray scale colormap of the {}``overlap matrices'' $O\left(t=60h,\tau=\left\{ 0h,3h,12h,24h,36h,48h\right\} \right)$
is presented in Figure 10. Black color of grayscale represents $O_{ij}=1$,
white color represents $O_{ij}=0.$ The most stable eigenvalue is
$\lambda_{492}.$%
\begin{figure}[h]
\begin{center}\includegraphics{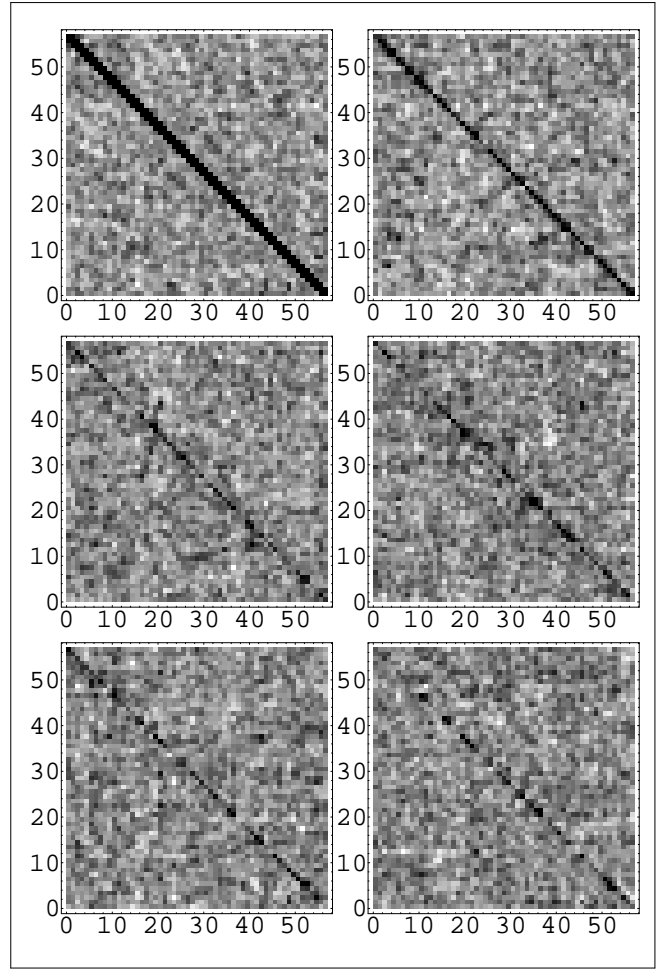}\end{center}

\caption{\label{10} The grayscale of overlap matrix $O\left(t,\tau\right)$
at $t=60h$ and $\tau=\left\{ 0h,3h,12h,24h,36h,48h\right\} $. }
\end{figure}
At lag $\tau=3$ hours the inter-VLAN interactions show the highest
degree of stability. For further lags the overall stability decays.
As the analysis of deviating eigenvectors content showed, the highly
interacting traffic time series are time series of service based VLANs,
intended for routing. Particular network services are evoked at the
same time and active for the same period of time, which explains the
stability and consequent decay of deviating eigenvectors of traffic
interactions.

\section{detecting anomalies of traffic interactions}

We assume that the health of inter-VLAN traffic is expressed by stability
of its interactions in time. Meanwhile, the temporal critical events
or anomalies will cause the temporal instabilities. The {}``deviating''
eigenvalues and eigenvectors provide us with stable in time snapshots
of interactions representative of the entire network. Therefore, these
eigenvectors judged on the basis of their IPR can serve as monitoring
parameters of the system stability.

Among the essential anomalous events of VLAN infrastructure we can
list violations in VLAN membership assignment, in address resolution
protocol, in VLAN trunking protocol, router misconfiguration. The
violation of membership assignment and router misconfiguration will
cause the changes in the picture of random and non-random interactions
of inter-VLAN traffic. To shed more light on the possibilities of
anomaly detection we conducted the experiments to establish spatial-temporal
traces of instabilities caused by artificial and temporal increase
of the correlation in normal non-congested inter-VLAN traffic. We
explored the possibility to distinguish different types of increased
temporal correlations. Finally, we observed the consequences of breaking
the interactions between time series, by injecting traffic counts
obtained from sample of random distribution. 

\textbf{\emph{Experiment 1}}

We selected the traffic counts time series representing the components
of the eigenvector which lies within the RMT bounds and temporarily
increased the correlation between these series for three hour period.
The proposed monitoring parameters show the dependence of system stability
on the number of temporarily correlated time series (see Figure 11).
Presented in Figure 11, left to right are (a) eigenvalue distribution
of interactions with two temporarily correlated time series, (b) IPR
of eigenvectors of interactions with two temporarily correlated time
series, (c) the overlap matrix of deviating eigenvectors with two
temporarily correlated time series. Top to bottom the layout shows
these monitoring parameters when correlation is temporarily increased
between 10 connections (d,e and f) and between 20 connections (g,h
and i). %
\begin{figure}[h]
\begin{center}\includegraphics{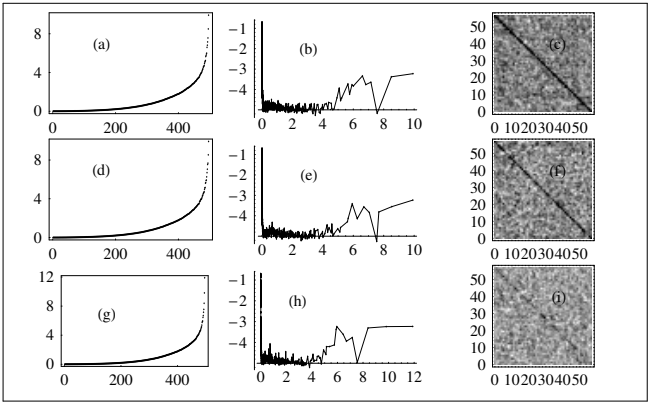}\end{center}

\caption{\label{11} Eigenvalues distribution, IPR and overlap matrix of deviating
eigenvectors. }
\end{figure}
One can conclude that increased temporal correlation between two time
series and between ten time series does not affect system stability.
Meanwhile, when the number of temporarily correlated time series reaches
the number of significant participants of $u^{497},$ which is calculated
as inverse of $I^{497}$and is equal to twenty two, the system becomes
visibly unstable. The largest eigenvalue changes from $10$ in stable
condition to $12$, the tail of inverse participation ratio plot is
extended and the diagonal of {}``overlap matrix'' disappears at
twenty temporarily correlated time series.

In Figures 12 (a, b, c and d ), the temporal correlation between ten
time series is traced with the matrix of sorted in decreasing order
of their components deviating from RMT eigenvectors.%
\begin{figure}[h]
\begin{center}\includegraphics{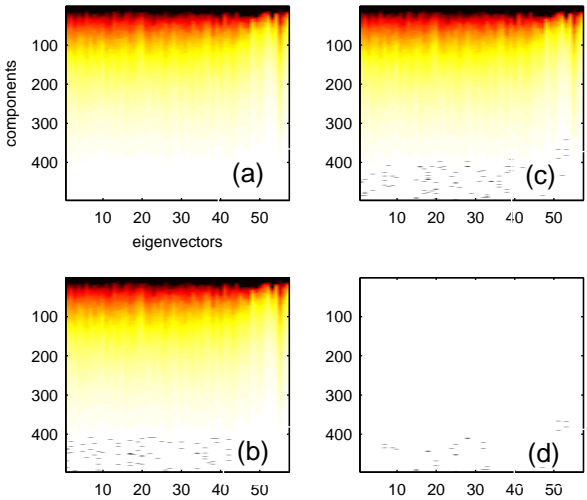}\end{center}

\caption{\label{12} Sorted deviating eigenvectors with injected correlation
among ten traffic time series.}
\end{figure}
The sorted in decreasing order deviating eigenvectors of $60h$ of
uninterrupted traffic are presented in Figure 12a. Then, after three
hours of uninterrupted traffic the weights of eigenvectors components,
which had zero value start changing, This is captured in Figure 12b.
Same process for traffic with induced three hours correlation is captured
in Figure 12c. The difference between results in Figures 12b and 12c
is presented in Figure 12d. The procedure used to visualize this produces
the high rate of false positive alarms. 

In addition, we visualize in Figure 13 the system instability during
temporal increase of correlation between twenty time series with spatial-temporal
representation of eigenvector $u^{497}$. %
\begin{figure}[h]
\begin{center}\includegraphics{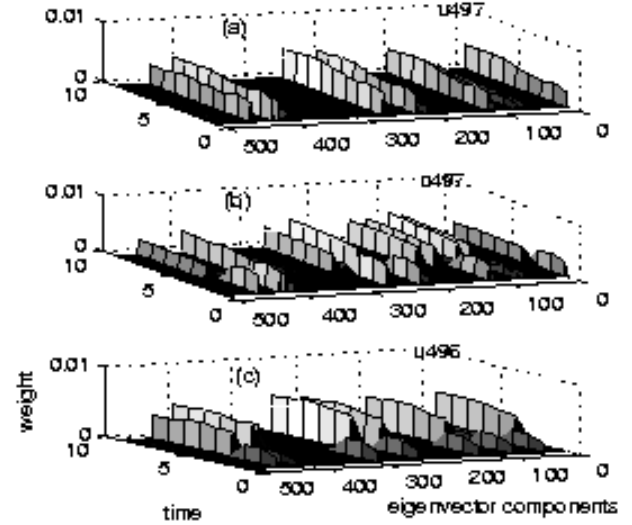}\end{center}

\caption{\label{13} (a) The weights of components of $u^{497}$ plotted for
time period from $36$ to $84$ hours of uninterrupted traffic with
6 hours interval. (b) The weights of components of $u^{497}$ plotted
with respect to the same time period, with induced three hours correlation.
(c) The weights of components of $u^{496}$ plotted with respect to
the same time period, with induced three hours correlation.}
\end{figure}
We used the weights of components of eigenvector $u^{497}$, defined
for IPR computation and plotted them with respect to time $t+\tau$,
where $t=36$ hours and $\tau=6n,$ where $n\in\left\{ 0,1,\dots,7\right\} $.
In Figure 13a the spatial-temporal pattern of $u^{497}$ captures
precise locations of system-specific interactions of uninterrupted
traffic for $84$ hours of observation. The abrupt change of this
pattern in Figure 13b indicates the starting point of induced correlation
between twenty traffic time series usually interacting in a random
fashion. It turns out, that the {}``normal'' stable pattern of eigenvector
$u^{497}$ moves to eigenvector $u^{496}$, when the interruption
ends. Thus, we are able to observe the end point of the induced correlations
in Figure 13c, which represents weights of components of eigenvector
$u^{496}$ plotted with respect to the same time intervals. With this
setup we are able to locate the anomaly in time and space. Translated
to network topological representation, the behavior of eigenvectors
$u^{497}$and $u^{496}$ during our manipulations with inter-VLAN
traffic may be monitored with the following graphs (see Figure 14).%
\begin{figure}[h]
\begin{center}\includegraphics{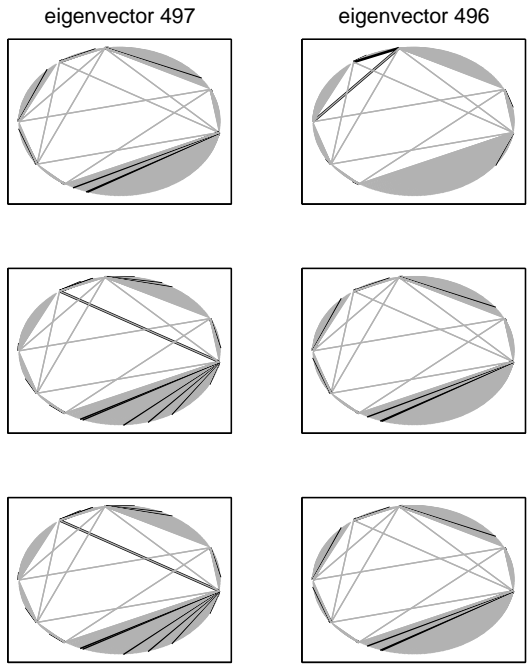}\end{center}

\caption{\label{14} Left column - behavior of $u^{497}$ during time period
from $48$h to $60$h with $6$h time window, induced correlation
starts at $54$h and lasts for $3$h. Right column - behavior of $u^{496}$
in same conditions.}
\end{figure}

\textbf{\emph{Experiment 2}}

In the previous experiment we injected just one type of increased
correlation among time series. Now we make two and three different
types of induced correlations produce different spatial-temporal patterns
on eigenvector $u^{497}$ components (see Figure 15). Time series
for temporal increase of correlation are obtained in the same way
as in Experiment 1. We temporarily increased the correlation between
series by inducing elements from distributions of sine function and
quadratic function, respectively for three hours.%
\begin{figure}[h]
\begin{center}\includegraphics{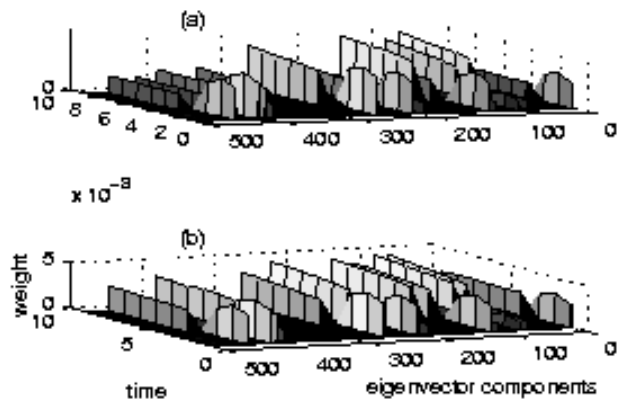}\end{center}

\caption{\label{15} (a) The weights of components of $u^{497}$ plotted for
time period from $36$ to $84$ hours with $6$ hours interval, two
different types of induced correlations. (b) The weights of components
of $u^{497}$ plotted with respect to the same time period, three
different types of induced correlations. }
\end{figure}
In Figure 15a, one type of three hours correlation is induced among
ten traffic time series and another type of correlation among other
ten time series. Three different types of three hours correlations
are induced among twenty traffic time series in Figure 15b. The sorted
in decreasing order content of significant components shows that time
series tend to group according to the type of correlation they are
involved in. 

\textbf{\emph{Experiment 3}}

Next we turn our attention to disruption of normal picture of inter-VLAN
traffic interactions. This can be done by injecting the traffic from
random distribution to non-randomly interacting time series for three
hours. We demonstrate it by examining the eigenvalue distribution,
the IPR and the deviating eigenvectors overlap matrix plotted in Figure
16.%
\begin{figure}[h]
\begin{center}\includegraphics{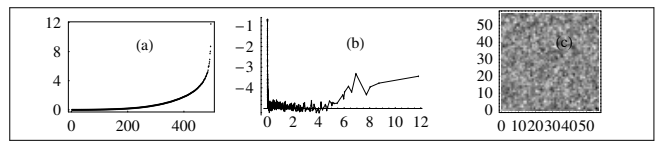}\end{center}

\caption{\label{16} Eigenvalues distribution, IPR and overlap matrix of deviating
eigenvectors of inter-VLAN traffic cross-correlation matrix $C$. }
\end{figure}
After $60$ hours of uninterrupted traffic, we injected elements from
random distribution to significant participants of $u^{497}$ for
three hours. The largest eigenvalue increases, from $10$ to $12.$
Extended IPR tail shows the larger number of \emph{localized} eigenvectors
and we observe the dramatic break in deviating eigenvectors stability.

\section{conclusion and future work}

The RMT methodology we used in this paper enables us to analyze the
complex system behavior without the consideration of system constraints,
type and structure. Our goal was to investigate the characteristics
of day-to-day temporal dynamics of the system of interconnected routers
with VLAN subnets of the University of Louisville. The type and structure
of the system at hand suggests the natural interpretation of the RMT-like
behavior and the RMT deviating results. The time stable random interactions
signify the healthy, and free of congestion traffic. The time stable
non-random interactions provide us with information about large-scale
network-wide traffic interactions. The changes in the stable picture
of random and non-random interactions signify the temporal traffic
anomalies. 

In general, the fact of sharing the universal properties of the bulk
of eigenvalues spectrum of inter-VLAN traffic interactions with random
matrices opens a new venue in network-wide traffic modeling. As stated
in \cite{Guhr1}, in physical systems it is common to start with the
model of dynamics of the system. This way, one would model the traffic
time series interactions with the family of stochastic differential
equations \cite{Farmer,Cont}, which describe the {}``instantaneous''
traffic counts \begin{equation}
g_{i}\left(t\right)=\left(d/dt\right)lnT_{i}\left(t\right),\label{eq22}\end{equation}
as a random walk with couplings. Then one would relate the revealed
interactions to the correlated {}``modes'' of the system.

Additional question that RMT findings raise in network-wide traffic
analysis is whether the found eigenvalues spectrum correlations and
\emph{localized} eigenvectors outside of RMT bulk can add to the explanation
of the fundamental property of the network traffic, such as self-similarity
\cite{Leland}. 

To summarize, we have tested the eigenvalues statistics of inter-VLAN
traffic cross-correlation matrix $C$ against the null hypothesis
of random correlation matrix. By separating the eigenvalues spectrum
correlations of random matrices that are present in this system, the
uncongested state of the network traffic is verified. We analyzed
the stable in time system-specific correlations. The analyzed eigenvalues
and eigenvectors deviating from the RMT showed the principal groups
of VLAN-router switches, groups of traffic time series interlinked
through the firewalls and groups of same service VLANs at every router.
With straightforward experiments on the traffic time series, we demonstrated
that eigenvalue distribution, IPR of eigenvectors, overlap matrix
and spatial-temporal patterns of deviating eigenvectors can monitor
the stability of inter-VLAN traffic interactions, detect and spot
in time and space of any network-wide changes in normal traffic time
series interactions.

As the reservation for the future work, we would like to investigate
the behavior of delayed traffic time series cross-correlation matrix
$C_{d}$ in the RMT terms. The importance of delay in measurement-based
analysis of Internet is emphasized in \cite{Zhang}. To understand
and quantify the effect of one time series on another at a later time,
one can calculate the delay correlation matrix, where the entries
are cross-correlation of one time series and another at a time delay
$\tau$ \cite{Mayya}. In addition, we are interested in testing the
fruitfulness of the RMT approach on the larger system of inter-domain
interactions, for instance, on 5-minute averaged traffic count time
series of underlying backbone circuits of Abilene backbone network.

\section*{acknowledgment}

This research was partially supported by a grant from the US Department
of Treasury through a subcontract from the University of Kentucky.
The authors thank Igor Rozhkov for consulting on the RMT methodology.
We thank Hans Fiedler, University of Louisville network manager, for
MRTG data of UofL routers system used in this study and helpful suggestions
in network interpretations of our results. We are grateful to Nathan
Johnson, University of Louisville super computing administrator, for
providing the computing environment and space.

\appendix

\section{RMT}

In this Appendix, we provide a short (and non-rigorous) explanation
of main concepts and glossary of terms used in the RMT studies. The
RMT approaches, which originated in nuclear and condensed matter physics
and later became common in many branches of mathematical physics \cite{Stockman},
have recently penetrated into econophysics, finance \cite{Bouchaud}
and network traffic analysis \cite{Barthelemy}. 

For the statistical description of complex physical systems, such
as, for example, atomic nucleus or acoustical reverberant structure,
the RMT serves as guiding light when one is interested in the degree
of mutual interaction of the constituents. As it turns out, the uncorrelated
energy levels or acoustic eigenfrequencies would produce qualitatively
different result from those obeying RMT-like correlations \cite{Stockman}.
Therefore, real (experimentally measured) spectra can help to decide
on the nature of interactions in the underlying system. To be specific,
ideally, symmetric system is expected to exhibit spectral properties
drastically different from the properties of generic one, and if the
spectral properties are those of RMT systems, other ideas of RMT can
be brought to the researcher aid.

To describe {}``awareness'' of the structural constituents about
each other, scientists in different fields use similar constructs.
Physicists use Hamiltonian matrix, engineers stiffness matrix, finance
and network analysts the equal-time cross-correlation matrix. Although
the physical meaning of mentioned operators can be different, the
eigenvalues/eigenvectors analysis seems to be a universally accepted
tool. The eigenvalues have direct connection to spectrum of physical
systems, while eigenvectors can be used for the description of excitation/signal/information
propagation inside the system. In physics, the RMT approaches come
about whenever the system of interest demonstrates certain qualitative
features in their spectral behavior. For example, if one looks at
nearest neighbor spacing distribution of eigenvalues and instead of
Poisson law\[
P\left(s\right)=\exp\left(-s\right),\]
 discovers {}``Wigner surmise''\[
P\left(s\right)=\frac{\pi}{2}s\exp\left(-\frac{\pi}{2}s^{2}\right),\]
one concludes (upon running several additional statistical tests)
that apparatus of RMT can be used for the system at hand, and system
matrix can be replaced by a matrix with random entries. For mathematical
convenience, these entries are given Gaussian weight. The only other
ingredient of this rather succinct phenomenological model is recognizing
the physical situation. For example, systems with and without magnetic
field and/or central symmetry are described by different matrix ensembles
(that is the set of matrices) with elements distributed within distribution
corresponding to the same $\beta$\[
P^{\left(\beta\right)}\left(H\right)\propto\textrm{exp}\left(-\frac{\beta}{4v^{2}}trH^{2}\right),\]
where the constant $v$ sets the length of the resulting eigenvalues
spectrum.

The very fact that RMT can be helpful in statistical description of
the broad range of systems suggests that these systems are analyzed
in a certain special \emph{universal} regime, in which physical or
other laws are undermined by equilibrated and ergodic evolution. In
most physical applications, a Hamiltonian matrix is rather sparse,
indicating lack of interaction between different subparts of the corresponding
object. However, if the universal regime is inferred from the above
mentioned statistical tests, it is very beneficial to replace this
single matrix with the ensemble of random matrices. Then, one can
proceed with statistical analysis using matrix ensemble for calculation
of statistical averages more relevant for the physical problem at
hand than the statistics of eigenvalues. The latter can be mean or
variance of the response to external or internal excitation.
\end{document}